\documentclass[aps,prb,reprint,citeautoscript,longbibliography,superscriptaddress,onecolumn,notitlepage]{revtex4-1}

\usepackage[colorlinks=true,allcolors=blue,breaklinks=true]{hyperref}

\usepackage{amsmath, amssymb}
\usepackage{graphicx, epsfig}
\usepackage{natbib}
\usepackage{textcomp} 
\usepackage{upgreek}
\usepackage{siunitx}
\usepackage{xcolor}

\newcommand{\appropto}{\mathrel{\vcenter{\offinterlineskip\halign{\hfil$##$\cr\propto\cr\noalign{\kern2pt}\sim\cr\noalign{\kern-2pt}}}}}

\begin{document}

\title{Time crystal optomechanics}

\author{J.~T.~M\"akinen}
\email[]{jere.makinen@aalto.fi}
\affiliation{Low Temperature Laboratory, Department of Applied Physics, Aalto University, FI-00076 Aalto, Finland}
\affiliation{QTF Centre of Excellence, Department of Applied Physics, Aalto University, FI-00076 Aalto, Finland}
\author{P.~J.~Heikkinen}
\affiliation{Low Temperature Laboratory, Department of Applied Physics, Aalto University, FI-00076 Aalto, Finland}
\affiliation{Department of Physics, Royal Holloway, University of London, Egham, Surrey, TW20 0EX, UK}
\author{S.~Autti}
\affiliation{Low Temperature Laboratory, Department of Applied Physics, Aalto University, FI-00076 Aalto, Finland}
\affiliation{Department of Physics, Lancaster University, Lancaster, LA1 4YB, UK}
\author{V.~V.~Zavjalov}
\affiliation{Low Temperature Laboratory, Department of Applied Physics, Aalto University, FI-00076 Aalto, Finland}
\affiliation{Department of Physics, Lancaster University, Lancaster, LA1 4YB, UK}
\author{V.~B.~Eltsov}
\affiliation{Low Temperature Laboratory, Department of Applied Physics, Aalto University, FI-00076 Aalto, Finland}

\date{\today}


\maketitle


{\bf Time crystals are an enigmatic phase of matter in which a quantum mechanical system displays repetitive, observable motion -- they spontaneously break the time translation symmetry. On the other hand optomechanical systems, where mechanical and optical degrees of freedom are coupled, are well established and enable a range of applications and measurements with unparalleled precision.
Here, we connect a time crystal formed of magnetic quasiparticles, magnons, to a mechanical resonator, a gravity wave mode on a nearby liquid surface, and show that their joint dynamics evolves as a cavity optomechanical system.
Our results pave way for exploiting the spontaneous coherence of time crystals in an optomechanical setting and remove the experimental barrier between time crystals and other phases of condensed matter.
}


The concept of time crystals comes very close to a perpetual motion machine\cite{PhysRevLett.109.160401, Sacha_2017, PhysRevLett.111.070402}, which is why experimental realizations of time crystals are never in true equilibrium; instead, they are driven or consist of quasiparticles that have a finite lifetime. Such time crystals have been created in a range of physical systems \cite{Mi_2021,doi:10.1126/sciadv.adg7541,zhang2017observation,giergiel2020creating, PhysRevLett.121.185301,PhysRevB.100.020406,Greilich_2024, 10.1063/5.0189649,PhysRevLett.120.215301, TimeCrystalJosephson, autti2022nonlinear} and coupled to other time crystals \cite{TimeCrystalJosephson, autti2022nonlinear}, but always in isolation from their environment. In particular, their inherent long-term coherence \cite{Greilich_2024} and versatility \cite{giergiel2024time} is yet to be investigated with coupled external degrees of freedom. Here, we realize controlled interaction between a continuous time crystal and a well-defined mechanical degree of freedom. We find that the coupled dynamics of two periodic processes, internal motion in the time crystal and the mechanical oscillations, form an optomechanics-like system \cite{RevModPhys.86.1391, PhysRevLett.122.153601}. Optomechanics comes with rich and well understood phenomenology that enables analysing the coupling and its effects on the time crystal in detail. We utilize a time crystal formed of magnetic quasiparticles of superfluid $^3$He, magnons, interacting with a nearby liquid surface. We show that the time crystal frequency is modulated by the motion of the free surface, providing a key component of a cavity optomechanical system. Additionally, we find that the coupling is nonlinear, enabling access to different regimes of optomechanics.

\subsection*{The optomechanical system}

\begin{figure}
  \centering \includegraphics[width=0.5\linewidth]{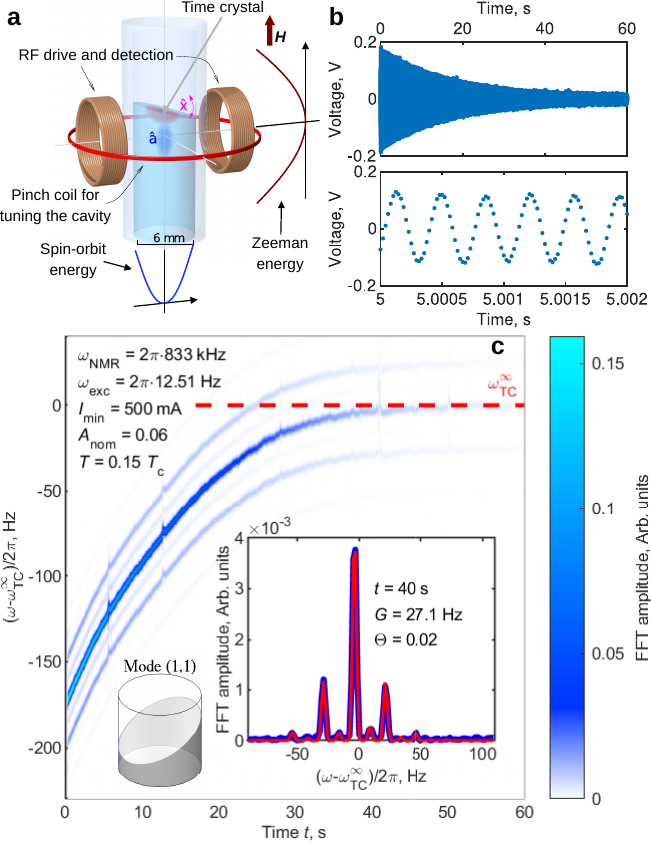}
  \caption{\label{fig:setupfig} {\bf Time crystal optomechanics.} {\bf a} The time crystals are formed of magnons (represented by operator $\hat{a}$) that are spatially trapped by the combined effect of the spin-orbit energy related to order parameter distribution (radial direction) and Zeeman energy controlled by the magnetic field profile (axial direction). The magnetic field profile ${\bf H}$ is used to move the time crystal against the free surface (red blob inside the container) or within the bulk liquid (blue blob inside the container). The time crystals couple to externally driven surface wave mode (represented by the position operator $\hat{x}$).
  {\bf b} The precession of magnetisation within the time crystal is observed as an induced, decaying sinusoidal voltage in the pick-up coils. 
  {\bf c} A sliding windowed fast Fourier transform (FFT) of the signal with on-resonance mechanical forcing reveals that the time crystal signal is accompanied by sidebands. The sidebands result from the frequency modulation of the time crystal signal, caused by the motion of the free surface. The red dash line indicates the mean frequency of the time crystal in the limit of vanishing magnon number $\omega_{\rm TC}^\infty = \langle \omega_{\rm TC} (t \rightarrow \infty) \rangle$. The measurements of the optomechanical coupling are carried out in this limit.  
  The large inset shows a snapshot of the signal including the fitted spectrum shape and parameter values in the region where the time crystal's frequency has ceased changing. 
  The fundamental surface wave mode driven in this Article is depicted in the small inset.}
\end{figure}

In our experiments, magnons are trapped in superfluid bulk by the combined effect of the order parameter distribution of the superfluid and a magnetic field profile, as illustrated in Fig.~\ref{fig:setupfig} {\bf a}. We pump magnons into this system using a short radio frequency (RF) pulse with frequency $\omega_{\rm NMR} \approx2\pi \times 833\,$kHz. After the pump is switched off, the magnons condense to the trap and spontaneously form a continuous time crystal \cite{10.1063/5.0189649} characterized by uniform precession of magnetisation of all condensed magnons. The precession frequency is the eigenfrequency of the time crystal wave function close to $\omega_{\rm NMR}$. The precession induces an oscillating voltage in the RF coils as shown in Fig.~\ref{fig:setupfig} {\bf b}. We use this signal for inferring the time crystal dynamics. While the magnons are non-equilibrium quasiparticles and slowly decay, the continuous time crystal preserves its coherent state for up to a few minutes, or about $10^8$ cycles.

In the paradigmatic optomechanical system the optical cavity is formed of two mirrors, one of them fixed and the other attached to a spring, allowing the mirror to move. In this setting, the cavity resonance frequency changes linearly with the position of the mirror, and radiation pressure couples the optical and mechanical degrees of freedom. Here, the time crystal is located in the vicinity of a free surface of the superfluid, where gravity waves of the surface make a mechanical oscillator. For small gravity wave amplitudes the surface is tilted uniformly and oscillates back and forth. The surface motion modifies the superfluid order parameter distribution and therefore the time crystal frequency. We can write the time crystal frequency as \cite{methods}
\begin{equation} \label{eq:CavityFreq}
 \omega_{\rm TC}(t) = \omega_0 + 2 \pi g  \times (\theta(t) - \theta_0)^2 \,,
\end{equation}
where $\omega_0$ is the time crystal's frequency when the free surface is orthogonal to the container axis, $g$ is an optomechanical coupling constant, $t$ is time, and $\theta(t) - \theta_0$ is the angle between the container axis and the surface normal. We have allowed for a small asymmetry $\theta_0$ describing the tilt of the container axis relative to gravity in the absence of surface motion. For small-amplitude surface waves close to the container axis $\theta(t, \omega_{\rm exc}) = \theta_{\rm max}(\omega_{\rm exc}) \sin (\omega_{\rm exc} t)$, where $\omega_{\rm exc}$ is the drive frequency and $\theta_{\rm max}(\omega_{\rm exc})$ characterises the frequency-dependent surface oscillation amplitude. We note that Eq.~\eqref{eq:CavityFreq} allows tuning between quadratic ($\theta_0 \ll \theta_{\rm max}$) and linear ($\theta_0 \gg \theta_{\rm max}$) optomechanics by using $\theta_0$ as a control parameter, providing access to dispersive optomechanics \cite{DispersiveOM}.

We control the position of the time crystal magnetically using a pinch coil. The time crystal can be positioned so that it is in the direct vicinity of the free surface, or in the bulk of the superfluid a few mm below the surface\cite{TimeCrystalJosephson,autti2022nonlinear}. In these two cases, the physical mechanism behind the coupling $g$ of the time crystal to the moving free surface is qualitatively different as discussed later in the Article. In all other respects the location makes no difference to the interpretation of the data.

Here, we drive the mechanical mode by moving the sample container nearly horizontally. The recorded time crystal signal, converted to frequency domain with windowed Fourier transform in Figure \ref{fig:setupfig} {\bf c}, exhibits sidebands when driven. These sidebands result from the optomechanical frequency modulation of $\omega_{\rm TC}$, caused by the motion of the superfluid free surface. A reconstruction of the experimental signal \cite{methods} (inset in Fig.~\ref{fig:setupfig} {\bf c}) allows extracting the products $G \equiv g \theta_{\rm max}^2$ and $\Theta \equiv \theta_0 \theta_{\rm max}^{-1}$. Note that the time crystal frequency is also increasing slowly across half a minute by about 150\,Hz before becoming stable. This is because local magnon density modifies the order parameter part of the trap\cite{self_trap} and the magnon number slowly decreases during the experiment due to dissipation\cite{magrelax_JLTP175}. We note that because of the very long life time of the time crystal and the changing frequency, the time crystal line width is determined by the length of the Fourier time window, not the time crystal quality factor. As such, the sideband width is not a measure of the dissipation of the mechanical mode, unlike in paradigmatic optomechanical systems \cite{RevModPhys.86.1391}.

\subsection*{Characterization of the mechanical mode}

\begin{figure}
  \centering \includegraphics[width=0.5\linewidth]{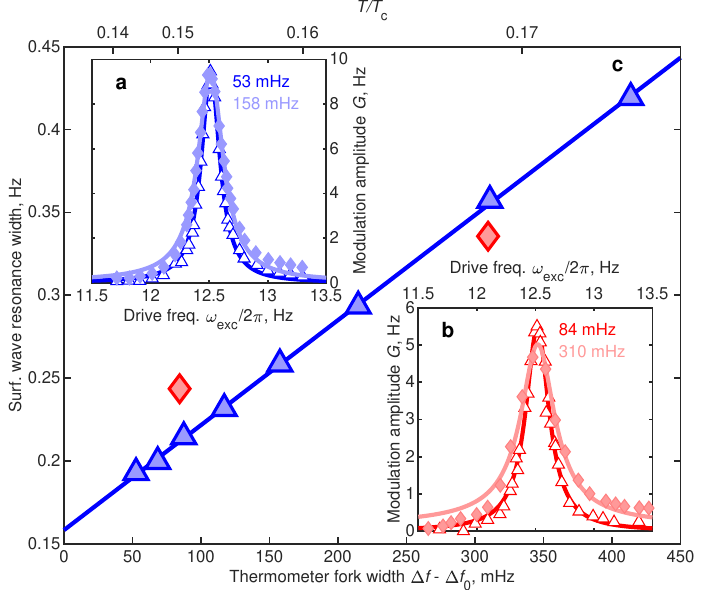}
  \caption{\label{fig:tempdep} {\bf The mechanical mode.} {\bf a} The frequency modulation amplitude  $G = g \theta_{\rm max}^2$ of the bulk time crystal measured as a function of the mechanical forcing frequency $\omega_{\rm exc}$. The solid lines are fits to a driven and damped harmonic oscillator response, from which the resonance frequency and width of the mechanical surface wave mode are determined. The thermometer fork width $\Delta f$ relative to the intrinsic width $\Delta f_0$ of the device for the two data sets is given in the legend. In the shown temperature interval the on-resonance amplitude $G$ remains approximately constant despite exponentially increasing resonance width, suggesting that coupling increases with the same exponent as a function of temperature. Here, unlike other measurements in this Article, the magnon number was kept constant by applying continuous pumping to an excited state in the confining trap\cite{PhysRevLett.120.215301}, as extremely long lifetime of bulk time crystal makes pulsed spectral measurements impractical. {\bf b} Data measured for the surface time crystal are extracted from the end of the decay after an RF excitation pulse, where the time crystal's frequency has stopped changing. 
  {\bf c} The extracted width of the surface wave resonance (bulk: blue triangles, surface: red diamonds) scales linearly with the the thermometer fork resonance width (solid line). This confirms that the damping of the mechanical mode that is coupled to the time crystal (mainly) originates from scattering of thermal excitations in the superfluid. The y axis intercept corresponds to the mechanical mode dissipation in the absence of superfluid thermal excitations, which may be caused by friction on the container walls, or quasiparticle states bound on the free surface\cite{forstner2024dynamicinteractionchiralcurrents}.
  }
\end{figure}

We use the time crystal to characterise the mechanical mode by sweeping the frequency of the mechanical forcing across the surface gravity wave resonance and recording the modulation amplitude of the time crystal frequency. The response is centred at $\approx 12.5\,$Hz (Fig.~\ref{fig:tempdep} {\bf a}, {\bf b}), which agrees well with the theoretical expectation for surface waves in this geometry within $0.1\,$Hz \cite{methods}.

The width of the surface wave resonance (its dissipation) is found to increase linearly with the quasiparticle density in the superfluid  (Fig.~\ref{fig:tempdep}). This observation connects the time crystal to the motion of the free surface beyond doubt, since the mechanical dissipation in the superfluid is caused by scattering of thermal excitations from the moving surface \cite{sw_yki}. The fermionic quasiparticle density can be measured separately by a tuning fork resonator, whose resonance width at low temperatures is linearly proportional to the quasiparticle density \cite{forktherm}, which in turn depends exponentially on temperature. Based on the matching resonance frequency and the observed linear dependence between the surface wave resonance width and that of the tuning fork's, we conclude that the observed resonance is connected with motion of the superfluid surface. 

The dissipation of the free surface oscillation releases heat into the superfluid, the amount of which depends on the drive power. We can measure the resulting temperature gradient in the superfluid by using the time crystal as a thermometer\cite{thermom_JLTP175} and comparing this with the reading of a separate tuning fork thermometer at the bottom of the sample container cylinder. The typical temperature increase at the free surface is a few µK, which corresponds to a few pW of heating power \cite{HMartin_thesis}. The extracted heating is then converted into the maximum surface tilt angle $\theta_\mathrm{max}$, which provides an independent measurement of the amplitude of the surface oscillations \cite{methods}.

\subsection*{Determining the nature of optomechanical coupling}

\begin{figure}
  \includegraphics[width=\linewidth]{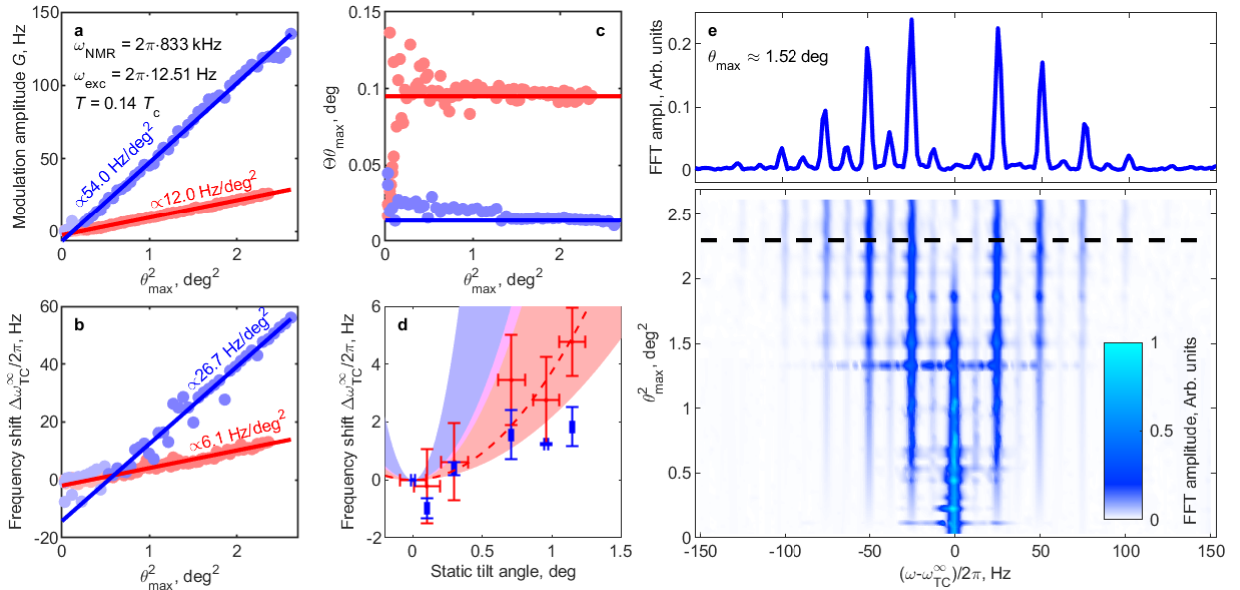}
  \caption{\label{fig:Coupling} {\bf The optomechanical coupling mechanisms.} {\bf a} With on-resonance mechanical forcing, the fitted time crystal frequency modulation amplitude $G$ (bulk: red points, surface: blue points) is found to depend linearly on $\theta_\mathrm{max}^2$ (fitted lines). {\bf b} The mean frequency shift $\Delta \omega_{\rm TC}^\infty = (\omega_{\rm TC}^\infty(\theta_{\rm max}) - \omega_{\rm TC}^\infty(\theta_{\rm max} = 0))$ is half of the frequency modulation amplitude, in agreement with Eq.~(\ref{eq:CavityFreq}). {\bf c} The product of the fitted asymmetry $\Theta$ and $\theta_{\rm max}$ (points) is found to be independent of the mechanical motion amplitude, as expected if $\theta_0$ originates from misalignment of the surface normal and container axis. Solid horizontal lines are a guide to the eye. {\bf d} As a function of the static tilt tilt angle, the surface time crystal frequency shift (red points) is consistent with the coupling constant $g_{\rm surf} = (2.2\div12.0)$\,Hz\,deg$^{-2}$ determined from dynamic measurements (red shaded region). The static shift for the bulk time crystal (blue points) is smaller than expected from the dynamic measurements (blue shaded region) $g_{\rm bulk} = (9.8\div54.0)\,$\,Hz\,deg$^{-2}$, showing that the optomechanical coupling is enhanced significantly in the dynamic case. The red dash line is a fit to points with $g_{\rm surf}=3.74$\,Hz\,deg$^{-2}$. The horizontal error bars correspond to the value of the static tilt from panel {\bf c} and the vertical error bars correspond to one standard deviation for measurements performed at different minimum coil currents ($N_{\rm surface} = 8$, $N_{\rm bulk}=3$).
  {\bf e} {\it Bottom}: Larger mechanical forcing amplitude results in larger frequency modulation, reflected in the number and amplitude of the side bands seen in the Fourier spectrogram of the time crystal signal. {\it Top}: For the largest free surface motion amplitudes the time crystal frequency modulation becomes large enough to completely diminish the central frequency band corresponding to $\omega_{\rm TC}^\infty$. The black dashed line in the bottom panel shows where this individual frequency spectrum lies in the plot below. 
  }
\end{figure}

With direct access to the mechanical mode amplitude, we now compare the coupling in Eq.~(\ref{eq:CavityFreq}) with observations. Figure \ref{fig:Coupling} {\bf a} shows that the measured time crystal frequency modulation amplitude $G$ is proportional to the square of the measured mechanical mode amplitude $\theta_\mathrm{max}^2$. For sinusoidal mechanical oscillations coupled quadratically to the time crystal as in Eq.~(\ref{eq:CavityFreq}), the mean frequency of the time crystal is expected to be shifted by half of the modulation amplitude. This is confirmed by Fig.~\ref{fig:Coupling} {\bf b}. Increasing the amplitude of the mechanical forcing leads to increased amplitude of the frequency modulation of $\omega_{\rm TC}$, seen as higher order side bands appearing and eventually the central band disappearing as shown in  Fig.~\ref{fig:Coupling} {\bf e}. In line with our interpretation of $\theta_0$, panel {\bf c} in Figure \ref{fig:Coupling} shows that the product of the fitted asymmetry $\Theta$ and $\theta_{\rm max}$, i.e. $\Theta \times \theta_{\rm max}$, for both the bulk and surface time crystals, is independent of the mechanical mode amplitude. Together these observations confirm that the coupling follows Eq.~(\ref{eq:CavityFreq}).

Finally, we can identify different contributions to the optomechanical coupling comparing the bulk and surface time crystal responses to applied static tilt of the free surface. Within experimental uncertainties, the surface time crystal frequency shift, and thus coupling $g_{\rm surf}$, is the same whether the tilt is static or dynamic (Fig.~\ref{fig:Coupling} {\bf d}). This implies that the surface time crystal frequency follows quasi-static changes in the confining trap as determined by the motion of the surface. The bulk time crystal $g_{\rm bulk}$ is smaller than the surface coupling if only static tilt is applied. This is reasonable as the effect of the tilted surface, via the order parameter part of the trapping potential, is felt less further away from the surface. 

With dynamic tilt, the bulk coupling is enhanced by more than an order of magnitude. Additionally, this enhancement increases with temperature, as can be seen in Fig.~\ref{fig:tempdep} {\bf a}, where larger mechanical line width at higher temperature does not reduce the measured signal in resonance. The additional coupling is caused by the superfluid flow linked to the surface displacement. This flow modifies the order parameter spatial distribution, and the magnitude of the effect depends exponentially on temperature \cite{de2011textures,thune_hydr}. The surface time crystal does not react to the flow, because close to the surface the boundary condition rigidifies the order parameter trap \cite{TimeCrystalJosephson,autti2022nonlinear} and thus the superflow is not able to change the trap shape. Postulating coincidence of dynamic and static coupling for the surface time crystal, we can estimate that at the lowest measured temperatures about 70\% of the the heat dissipated by the moving free surface is carried away by surface-bound Andreev states\cite{autti2023transport}. Theoretical details and a phenomenological derivation supporting the above findings can be found in \cite{methods}. We also note that the measured asymmetry of the coupling in Fig.~\ref{fig:Coupling} {\bf c} is smaller for the bulk time crystal than for the surface one. This has natural explanation as the residual static tilt of the container axis with respect to gravity becomes less important for the bulk time crystal, where the coupling is mostly dynamic, due to superflow.

\subsection*{Discussion and Outlook}

Based on the observations reported in this Article, we conclude that continuous time crystals can be coupled to a macroscopic mechanical oscillator and that the resulting coupled dynamics are described by an optomechanical Hamiltonian, thus combining the inherent coherence of time crystals with the sensitivity of optomechanical systems. In our present system the optomechanical coupling is found to be predominantly quadratic \cite{PhysRevA.99.053801}, but is expected to be tuneable all the way to linear by adjusting $\theta_0$. The underlying idea, tuning the time crystal period by coupling to an external degree of freedom, is very general and we expect that time crystal optomechanics can be realised in other physical systems. This will open new avenues for research on time crystals as optomechanical systems have shown the capacity to push the boundaries of both fundamental and applied physics \cite{Optomechanics4QT}, for example via measuring weak forces such as gravitational waves \cite{doi:10.1126/science.256.5055.325, TAccadia_2012, 10.1093/ptep/ptaa125}, and cooling mechanical degrees of freedom down to their ground state \cite{Piotrowski_2023} and even entangling them \cite{doi:10.1126/science.abf5389}.

We also note that the frequency modulation scheme in our experiments is similar to frequency comb generation using a modulated continuous wave laser \cite{Fortier_2019}, paving way for utilizing time crystals for precision spectroscopy \cite{Lesko_2021}. For suitably chosen large mechanical drive amplitudes the system becomes transparent at the time crystal's central frequency, Fig.~\ref{fig:Coupling} {\bf e}. Control over the transparency of a truly macroscopic and long-living quantum state may prove useful for example for quantum state storage \cite{Saglamyurek_2021}, as the system additionally has continuously tunable central frequency (via magnetic field) and band separation (via cylinder radius or coupling to another type of mechanical resonator). We also note that the underlying physical system is a topological superfluid, providing wide possibilities from dark matter research \cite{PhysRevLett.129.211801,PhysRevD.110.043005} to detection of topological defects \cite{M_kinen_2019, PhysRevLett.86.268} using the optomechanical time crystal system as an instrument. 

Finally, it is interesting to consider whether time crystal optomechanics could be realized in the quantum regime. This requires reducing the mass of the mechanical mode, and increasing its resonance frequency and quality factor. These enhancements can be achieved by e.g. by using nanoelectromechanical resonators as the mechanical degree of freedom \cite{PhysRevB.107.014502}. Besides their significantly lower mass, such devices can also be designed to have orders of magnitude larger resonance frequencies and quality factors, both of which are essential for reaching the quantum regime. Potentially, such a setup allows for mechanical resonance frequencies matching or even exceeding the optical mode frequency, yielding access to the mechanical dynamical Casimir effect \cite{jiang2024realizingmechanicaldynamicalcasimir, PhysRevLett.77.615} and yet unexplored regimes of optomechanics. Moreover, an attached micromagnet could be utilized to enhance coupling with the magnonic time crystal. We emphasise that similar research avenues may be accessible also at room temperature by utilizing suspended yttrium-iron-garnet (YIG) film resonators \cite{doi:10.7566/JPSJ.89.113702}, which can host magnon time crystals at room temperature \cite{PhysRevB.104.214416, Borisenko_2020}. Thus, our work paves way for utilising time crystals as tools for research, and as an integral part of hybrid systems for quantum technology.


\renewcommand{\theequation}{M\arabic{equation}}
\setcounter{figure}{0}
\setcounter{table}{0}
\setcounter{equation}{0}

\clearpage

\section*{Methods}

\subsection*{Experimental setup}

The sample container we use is a cylindrical quartz-glass tube (15~cm long, $2R=5.85$\,mm diameter, Fig.~\ref{fig:setupfig}). The $^3$He placed in the cylinder is cooled down by a nuclear demagnetisation refrigerator into the superfluid B phase. The lower end of the sample container connects to a volume of sintered silver powder surfaces, thermally linked to the nuclear refrigerant. This allows cooling the $^3$He in the cylinder down to 130~µK. 

Temperature of the superfluid is measured using a quartz tuning fork\cite{forktherm,2008_forks}, immersed in the superfluid.  In the low-temperature regime investigated in this manuscript, the fork's resonance width depends linearly on the thermal excitation density, which in turn depends on temperature as $\propto \exp (- \Delta / k_{\rm B} T)$, where $\Delta$ is the superfluid gap and $k_{\rm B}$ is the Boltzmann constant. Additionally, we can utilize the relaxation rate of the bulk time crystal as a thermometer\cite{thermom_JLTP175} -- its relaxation rate likewise depends exponentially on temperature, $\tau^{-1} \propto \exp (- \Delta / k_{\rm B} T)$.

The pressure of the superfluid sample is equal to saturated vapour pressure, which is vanishingly small at these low temperatures. The superfluid transition temperature at this pressure is $T_\mathrm{c}\approx\, $0.93~mK. The transverse nuclear magnetic resonance (NMR) pick-up coil, placed around the sample container, is part of a tank circuit resonator with $Q\approx150$. The setup includes also a pinch coil to create a minimum along the vertical axis of the otherwise homogeneous axial magnetic field. The resonance frequency of the tank circuit is 833~kHz in all measurements presented in this paper, 
corresponding to an external magnetic field of 25~mT. We use a cold preamplifier\cite{pjheikki_thesis} and room temperature amplifiers to amplify the voltage induced in the NMR coils. 

The free surface is positioned 3~mm above the location of the magnetic field minimum. We adjust the liquid level by removing $^3$He starting from the originally fully filled sample container while measuring the pressure of $^3$He gas in a calibrated volume.

The time crystals are created by a short ($\sim 1\,$ms) excitation pulse via the NMR coils. The pulse is resonant with an excited state in the trap that confines the magnons (details of the trap are described below). In about 0.3\,s, well after the end of the pump pulse, the pumped magnons spontaneously form a time crystal on the ground state of the trap in a Bose-Eistein condensate phase transition. The time crystal frequency and magnon number can be inferred from the AC voltage induced in the NMR coils, originating from the coherent precession of magnetisation in the time crystal at $\omega_\mathrm{TC}$. Details of the time crystal wave functions and signal readout are explained in Ref.~\citenum{10.1063/5.0189649} and the Methods sections of Refs.~\citenum{TimeCrystalJosephson,autti2022nonlinear}.

\subsection*{Signal analysis}

The precession of magnetization in the time crystal produces a voltage signal that is picked up by pick-up coils. This signal is fed to a lock-in amplifier, which is locked a few kHz above the time crystal's precession frequency and is thus used for the frequency downconversion. The lock-in output is sampled at 48 kHz frequency. We transform the wave record to frequency space by a windowed fast Fourier transformation (FFT) with a $3 \times 10^4$ point window size and a $10\%$ shift between windows.

We then trace the frequency of the central band in the FFT signal in time. In most cases this is the maximum of the FFT signal, but care should be taken at large mechanical drive amplitudes, where the sideband amplitude may exceed the central band amplitude. This trace is then used by the fitting algorithm.

The amplitude of the FFT spectrum is fit separately in each window to the FFT of the model signal
$U(t) = A \sin \int_0^t \omega_{\rm TC}(t')\,dt'$ with $\omega_{\rm TC}$ from Eq.~(\ref{eq:CavityFreq}) and $\theta(t)=\theta_{\rm max} \sin \omega_{\rm exc} t$. The surface modulation frequency $\omega_{\rm exc}$ is determined from the FFT of the geophone record (see the next section), measured simultaneously with the time crystal signal. There are four fitting parameters: the overall amplitude $A$, combinations $G = g \theta_{\rm max}^2$ and $\Theta = \theta_0 \theta_{\rm max}^{-1}$ describing coupling to the surface mode and the average frequency $\langle\omega_{\rm TC}\rangle$. Note that it is important for the proper fit to allow $\omega_0$ in Eq.~(\ref{eq:CavityFreq}) to depend on time as magnons are decaying from the trap, since the widths of the spectral bands are affected by this time dependence, especially near the beginning of the signal. From a single fitting parameter $\langle\omega_{\rm TC}\rangle$ for a window, we model $\omega_0(t)$ using time dependence of the traced frequency described above. 

\subsection*{Mechanical forcing and calibration}

The cryostat is levitated on top of four active air spring dampers with attached distance sensors. These dampers are normally used to isolate the cryostat from mechanical vibrations in the surrounding support structures. The sensors feed their output to a proportional-integral-derivative (PID) controller. To induce mechanical forcing, we modulate the set point of one of the air springs as $A_{\rm nom} \sin(\omega_{\rm exc} t)$ resulting in small deviations of the cryostat's tilt angle with the drive frequency $\omega_{\rm exc}$. The air springs are located approximately two meters above the sample cell, resulting in nearly horizontal oscillations of the sample. Similar mechanical forcing has been applied previously for surface wave studies in superfluid $^3$He and $^4$He \cite{sw_yki}.

We calibrate the tilt angle by attaching a laser pointer to the cryostat and applying a static tilt by changing the setpoint of the airspring we use for modulation. We then monitor the position of the laser pointer at fixed distance and calculate the corresponding tilt angle. This way, we are able to apply tilt angles $\lesssim 1.2^\circ$. In dynamic measurements the applied setpoint shifts are kept well below this value. For recording amplitude of the dynamic drive, the cryostat is equipped with the geophone. It is installed at room temperature on the level of the air spring dampers and thus conversion of the geophone signal to the surface oscillation amplitude requires calibration described below.

We connect the nominal mechanical forcing amplitude $A_{\rm nom}$ to the motion of the cryostat using a voltage $V_{\rm gp}$ produced by the geophone. The result is shown in Extended Data Figure~1 (from the same data set as in Fig.~\ref{fig:Coupling} of the main text). We find that the mechanical motion amplitude is well described by a function of the form
\begin{equation}
 V_{\rm gp} = C \cdot A_{\rm nom}^\nu + B\,,
\end{equation}
which contains three fitting parameters: a scale factor $C$, free exponent $\nu$, and constant $B$ to allow for non-zero base level in the measured voltage. 

Converting the mechanical forcing as measured by the geophone to excitation of the free surface motion can be done in three steps. If we drive the surface mode on resonance, the applied drive power is equal to dissipated power. The dissipation is observed as a heat flow into the superfluid, originating from the free surface motion. The resulting temperature gradient can be measured directly by comparing the temperature measured by the thermometer fork at the bottom of the container and using the time crystal as a thermometer at the top of the container\cite{thermom_JLTP175}, see Extended Data Figure~2~{\bf a}.

The temperature gradient can be converted to power using the known thermal conductivity of superfluid $^3$He-B in a cylindrical container\cite{HMartin_thesis}. The thermal resistance of $^3$He-B at similar experimental conditions (pressure, temperature, magnetic field) was measured to be $R_{\rm T}' \approx 0.15$~µK\,pW$^{-1}$ for a 5-cm-long cylindrical container with 8$\,$mm diameter. Our cylindrical container has a $\approx$6$\,$mm diameter and the magnon condensate is located some $\approx 14\,$cm above the bottom volume containing the thermometer fork. Scaling the thermal resistance with the lengths and inverse square radii, we get an estimate $R_{\rm T} \approx 0.75$~µK\,pW$^{-1}$ between the magnon time crystal and the heat exchanger volume in our experimental geometry.

The power can be converted into a range of motion as follows. The fraction of energy lost per cycle in a damped harmonic oscillator is
\begin{equation}
    \frac{\Delta E}{E(\theta_\mathrm{max})} = 1- e^{-2 \pi /Q}
\end{equation}
where $Q$ is the quality factor of the oscillator. The quality factor estimated as the temperature dependent part of the dependence shown in Fig.~\ref{fig:tempdep} gives $Q \approx 375$, while the total quality factor, including the zero-temperature offset for the surface wave resonance width, is $Q \approx 65$. The gravitational potential energy stored in the free surface motion at the maximum amplitude of the oscillation cycle is $E(\theta_\mathrm{max})=(\pi/8) \rho_{\rm He} g_{\rm g} R^4 \theta_\mathrm{max}^2$, where $\rho_{\rm He}=81.9\,$kg\,m$^{-3}$ is the density of the superfluid \cite{Wheatley_RMP} and $g_{\rm g}=9.81\,$m\,s$^{-2}$ is the free-fall acceleration. Thus, the dissipation power becomes

\begin{equation}
    P = \frac{\omega}{16} \left( 1- e^{-2 \pi /Q} \right) \rho_{\rm He} g_{\rm g} R^4 \theta_\mathrm{max}^2
\end{equation}
which yields $P/\theta_\mathrm{max}^2 \approx 8.1\,$pW\,deg$^{-2}$ for the total quality factor $Q \approx 65$.

We can now express the measured temperature difference $\Delta \tilde{T}(A_{\rm exc}) = |\Delta T_{\rm TC} (A_{\rm exc}) - \Delta T_{\rm fork}$, where $A_{\rm exc} = V_{\rm gp}(A_{\rm nom})/V_{\rm gp}(A_{\rm nom} = 0.098)$ is the normalized amplitude, as
\begin{equation} \label{eq:tiltcalib}
    \frac{\Delta \tilde{T}}{\theta_{\rm max}^2} = \frac{R_{\rm T} P}{\theta_{\rm max}^2} \approx 6.1\,\mu{\rm K}\,{\rm deg}^{-2}\,.
\end{equation}
Thus, drive amplitude is determined as $\theta_{\rm max}^2 {\rm (deg)}^2 \approx \Delta \tilde{T}/6.1\,$µK. To obtain direct relation between $\theta_{\rm max}$ and $A_{\rm exc}$ (and thus $V_{\rm gp}$), we perform a single parameter fit shown in Extended Data Figure~2~{\bf b}, resulting in $\theta_{\rm max}^2{\rm (deg)}^2 \approx 2.62\,A_{\rm exc}$. In analysis and plotting of the data in the main text as a function of the drive amplitude $\theta_{\rm max}$, we convert applied excitation $A_{\rm nom}$ to $V_{\rm gp}$ and then to $\theta_{\rm max}$ using calibration expressions above. 

In Figure \ref{fig:Coupling} {\bf d} the upper edge of the shaded areas correspond to the calibration obtained as explained above, using the full $Q \approx 65$. This calibration is used for $\theta_{\rm max}$ scale in other panels of Fig.~\ref{fig:Coupling}. In general, we may expect some of the dissipated heat to be carried away by a layer of surface bound states on the walls of the container and the free surface \cite{autti2023transport,autti2020fundamental}. The temperature-independent part of dissipation is generated at the surfaces. If all of this heat is carried away along the surface, never entering the bulk, then $Q \approx 375$ will be appropriate to calibrate the tilt angles instead. The lower edge of the shaded regions in Fig.~\ref{fig:Coupling} {\bf d} correspond to this limit. It is plausible that only a part of the generated heat escapes the bulk and these two extreme limits we take as uncertainty of the calibration which determines uncertainty range of the determined optomechanical coupling $g$. 

If we postulate that for the surface time crystal static and dynamic coupling coincide, we can fit the dynamic coupling to the static measurements to pick up a particular heat release fraction from the uncertainty band. The fit shown by the dashed line in Fig.~\ref{fig:Coupling} {\bf d} gives $g_{\rm surf} \approx 3.74$~Hz\,deg$^{-2}$. This value corresponds to 84\% of the temperature-independent part of the dissipated heat being carried away by the surface bound states. From the total dissipation at the lowest experimental temperature this makes 69\%.

\subsection*{Gravity wave resonances}

For inviscid, incompressible, and irrotational flow, the dispersion of gravity waves in a cylindrical container much deeper than its diameter is given by \cite{SW_damping1957}
\begin{equation}
    \omega_{\rm SW}^2 = g_{\rm g} k_{i} \left( 1 + \frac{\sigma k_i^2}{g \rho_{\rm He}} \right)\,,
\end{equation}
where $k_{i}$ is the wave number, and $\sigma = 155\,$µN/m is the surface tension of $^3$He \cite{surftens}.

The velocity field ${\bf u}$ related to a scalar potential $\phi$ can be calculated as
\begin{equation}
 {\bf u} = - \nabla \phi\,.
\end{equation}
For an infinitely deep cylindrical container, the potential associated with the planar fundamental mode is \cite{PhysRevFluids.4.012801}
\begin{equation}
 \phi = A J_1 (k_1 r) e^{k_1 z} \sin(\omega t) \sin(\varphi)\,,
\end{equation}
where $A = \omega_{\rm m} \theta_{\rm max} k_1^{-2}$ is the wave amplitude, $J_i$ is the Bessel function of the first kind or order $i$, $k_1$ is the wave number of the fundamental mode, $z$ is the vertical coordinate ($z = 0$ at fluid surface, negative values towards the fluid), and $\varphi$ is the azimuthal angle.

The resulting velocity field is then 
\begin{equation} \label{eq:flowfield}
\begin{split}
\vec{\bf u} = & -\frac{A}{2}k_1 \left( J_0 (k_1 r) - J_2 (k_1 r) \right) e^{k_1 z} \sin (\omega t) \sin( \varphi) \hat{\bf r} \\
 & - \frac{A}{r} J_1 (k_1 r) e^{k_1 z} \sin (\omega t) \cos ( \varphi) \hat{\bf \varphi} \\
 & - A k_1 J_1 (k_1 r) e^{k_1 z} \sin (\omega t ) \sin ( \varphi) \hat{\bf z}\,,
 \end{split}
\end{equation}
which additionally sets the magnitude of $k_1$ by requiring the radial flow to be zero at the container wall. In other words, the wave number $k_1$ is the first solution to equation
\begin{equation}
 J_0 (k_1 R) - J_2 (k_1 R) = 0\,,
\end{equation}
setting $k_1 \approx 1.8412/R$.

Taking into account the meniscus effect, the gravity wave dispersion relation is modified to \cite{freq_menisc}
\begin{equation}
    \omega_{\rm SWm}^2 = \omega_{\rm SW}^2 \left( 1 - \frac{2 \sigma k_i}{g_{\rm g} \rho_{\rm He} R} \right) \,.
\end{equation}
For the lowest mode with $k_1$, we get $\omega_{\rm SWm}/2\pi \approx 12.4\,$Hz, in good agreement with experimental observations. 


\subsection*{Surface wave oscillations near the container axis}

The surface height profile of the first mode takes the form \cite{miles_1984}
\begin{equation} \label{eq:HeightProfile}
 h (r,\phi,t) = A J_{1}(k_1 r) \sin (\omega t) \sin \varphi\,,
\end{equation}
To derive the maximum tilt angle during the time evolution of the surface wave, we set $\sin \varphi = 1$. We can estimate the tilt angle $\theta$ from the radial derivative of Eq.~\eqref{eq:HeightProfile}:
\begin{equation}
 \tan \theta \equiv \frac{\partial h}{\partial r} = \frac{A k_1}{2} \left[ \left( J_0(k_1 r) - J_2(k_1 r) \right) \sin (\omega t) \right]\,.
\end{equation}
Solving for $\theta$ and taking the limit $r \rightarrow 0$ (the time crystal is located close to $r = 0$), to leading order we get
\begin{equation}
 \theta_{r \rightarrow 0} \approx \arctan \left[ \frac{A k_1}{2} \sin (\omega t) \right] \approx \frac{A k_1}{2} \sin (\omega t) \,,
\end{equation}
where the last approximation results from $Ak_1 /2 \ll 1$. Therefore, we arrive to
\begin{equation} \label{eq:DriveFreq}
 \theta(t) \approx \theta_{\rm max} \sin(\omega t)\,,
\end{equation}
where we have set $\theta_{\rm max} \equiv Ak_1 /2$. This is the time dependence used in the main text.

\subsection*{Magnon trapping potential}

The axial trapping potential for (optical) magnons in the superfluid is set by the magnetic field profile as
\begin{equation}
    U_{\parallel}(z)/\hbar = \omega_{\rm L} (z) = |\gamma| H(z),
\end{equation}
where $\omega_{\rm L}$ is the local Larmor frequency and $\gamma$ is the gyromagnetic ratio of $^3$He. The magnetic field profile is created with a solenoid creating a field with inhomogeneity on the level of $10^{-4}$, and by a pinch coil which produces a local minimum in the magnetic field along the vertical axis, shown in Fig.~\ref{fig:setupfig}.

The radial trapping potential is set by the textural configuration via the spin-orbit interaction
\begin{equation} \label{Eq:text_trap}
    U_{\perp} (r)/\hbar = \frac{4 \Omega_{\rm L}^2}{5 \omega_{\rm L}} \sin ^2 \frac{\beta_{\rm L}(r)}{2}\,,
\end{equation}
where $\Omega_{\rm L}$ is the B-phase Leggett frequency and $\beta_{\rm L}$ is the polar angle of the Cooper pair orbital angular momentum, measured from the direction of the static magnetic field. In the absence of gravity waves, the minimum energy configuration corresponds to the so-called flare-out texture \cite{PhysRevB.15.199}. For low magnon numbers, such as at the end of the time crystals' decay in our experiments, the flare-out texture with $\beta_{\rm L} \propto r$ near the axis results in an approximately harmonic trapping potential with a characteristic size set by the magnetic healing length $\xi_{\rm H}$ \cite{self_trap}. However, for large magnon numbers the potential is heavily modified and can even become self-bound \cite{QBall, 10.1063/5.0189649} or box-like \cite{self_trap}. The total trapping potential is given by
\begin{equation}\label{Eq:tot_trap}
    U_{\rm tot} ({\bf r}) = U_{\parallel}(z) + U_{\perp} (r)\,.
\end{equation}

For zero or low pinch coil currents the spatial variation of $U_{\parallel}$ is negligible, and the trapping potential is fully set by the textural part. The surface energy orients the orbital angular vector perpendicular to the surface, setting $\beta_{\rm L} = \pi/2$ at container walls and $\beta_{\rm L} = 0$ at the free surface. The spatial variation of $\beta_{\rm L}$ creates the surface trap, which can additionally be identified by the faster magnon relaxation rate \cite{TimeCrystalJosephson}.

\subsection*{Optomechanical Hamiltonian}

We describe the combination of a time crystal in the superlfuid trap and the moving free surface as an optomechanical system with the Hamiltonian 
\begin{equation} \label{eq:Hamiltonian}
\hat H = \hbar \tilde{\omega}_{\rm TC} \hat a^\dagger \hat a + \hbar \omega_{\rm m} \hat b^\dagger \hat b + 2 \pi \hbar g_1 \hat a^\dagger \hat a \left( \hat b^\dagger + \hat b \right) + 2 \pi \hbar g_2 \hat a^\dagger \hat a \left( \hat b^\dagger + \hat b \right)^2 + \mathrm{mech. drive} +\mathrm{damping},
\end{equation}
where $\hat a^\dagger$ and $\hat a$ are  magnon creation and annihilation operators, $\hat b^\dagger$ and $\hat b$ are mechanical mode quanta (here ripplons) creation and annihilation operators, $g_1$ and $g_2$ are the linear and quadratic coupling constants, respectively, and $\omega_{\rm m}$ corresponds to the resonance frequency of the fundamental mode. We follow approach of Ref.~\citenum{PhysRevA.95.053858} but the laser-drive-like terms are not relevant to our experiments and do not appear in Eq.~(\ref{eq:Hamiltonian}).

Eq.~\eqref{eq:CavityFreq} in the main text can be derived from Eq.~\eqref{eq:Hamiltonian} by the following substitutions: $g_2 \rightarrow g$, $g_1 \rightarrow 2 g \theta_0$, and $\tilde{\omega}_{\rm TC} \rightarrow \omega_{\rm 0} + 2 \pi g \theta_0^2$, where $\theta_0$ is an effective static tilt of the surface in the absence of oscillations. Thus, tilting the free surface with respect to the gravitational potential results in the time crystal frequency being modulated as
\begin{equation}
 \omega_{\rm TC}(t)= \omega_0 + g \times \left( \theta(t) - \theta_0 \right)^2\,,
\end{equation}
where $\theta_0$ corresponds to the global minimum energy and $\theta(t)$ is the instantaneous angle of the surface normal w.r.t. gravity. Note that by changing geometry (expressed as $\theta_0$), the coupling can be smoothly tuned from quadratic to predominantly linear.

For mechanical motion as given in Eq.~\eqref{eq:DriveFreq}, we get
\begin{align} \label{eq:FreqMod}
 \Delta \omega_{\rm TC} = & \, g \bigg[ -\frac{1}{2} \theta_{\rm max}^2 \cos(2\omega  t) \nonumber \\
 & \quad - 2 \theta_{\rm max} \theta_0 \sin (\omega t) + \theta_0^2 + \frac{1}{2}\theta_{\rm max}^2 \bigg] \,,
\end{align}
where $\Delta \omega_{\rm TC} = \omega_{\rm TC}(t) - \omega_0$. From Eq.~\eqref{eq:FreqMod} one can see that there are components at both $\omega$ and $2 \omega$. The average frequency reads
\begin{equation} \label{eq:average_freq}
    \left \langle \Delta \omega_{\rm TC} \right \rangle_t = g \left( \theta_0^2 + \frac{1}{2} \theta_{\rm max}^2 \right)\,,
\end{equation}
 that is, the mean frequency increases as $\propto \frac{1}{2} g \theta_{\rm max}^2$. The origin of $g$ from free-energy considerations in the superfluid is discussed in the supplemental material.


\clearpage 

\bibliography{bibliography}
\bibliographystyle{naturemag}


\section*{Acknowledgments}
We thank G.~E.~Volovik, M.~Krusius, M.~A.~Silaev, I.~A.~Todoshchenko, M.~S.~Manninen, J.~T.~Tuoriniemi, and L. Mercier de L\'epinay for stimulating discussions. The research was done using facilities of the Low Temperature Laboratory infrastructure supported by the Aalto University. We acknowledge the computational resources provided by the Aalto Science-IT project. The work is supported by the Academy of Finland (present name: Research Council of Finland) with Project No. 332964.  P.~J.~Heikkinen acknowledges financial support from the V\"{a}is\"{a}l\"{a} foundation of the Finnish Academy of Science and Letters and from the Finnish Cultural Foundation, and S.~Autti acknowledges financial support from the UK EPSRC (EP/W015730/1) and Jenny and Antti Wihuri Foundation.


\section*{Author contributions:}
The manuscript was written by JTM and SA, with contributions from PJH and VBE. Experiments were conducted by PJH, SA, VVZ, and VBE. Theoretical analysis was led by JTM, with contributions from PJH, SA, and VBE. JTM analysed the data and prepared the figures. The project was supervised by VBE. All authors discussed the results.

\section*{Competing interests:}
The authors declare no competing interests.

\section*{Materials and correspondence}
Correspondence and materials requests should be addressed to J.T.M.


\clearpage

\renewcommand{\thefigure}{\arabic{figure}} 
\setcounter{figure}{0}
\renewcommand{\figurename}{Extended Data Figure}

\section*{Extended Data}

\begin{figure}[!h]
  \centerline{\includegraphics[width=\linewidth]{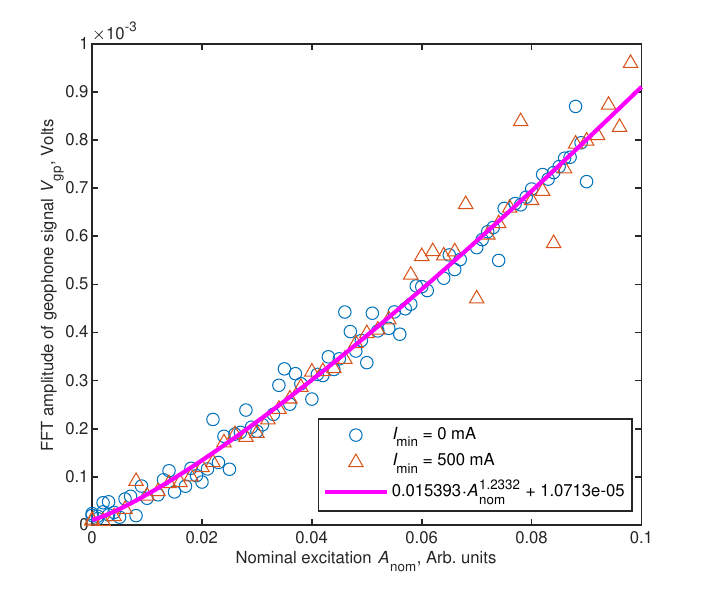}}
  \caption{\label{fig:ExcCalib} {\bf Calibration of the excitation amplitude} To reduce noise in determination of the mechanical excitation amplitude, we estimate the functional dependence of the real excitation amplitude, measured by the geophone voltage $V_{\rm gp}$, on the nominal excitation amplitude $A_{\rm nom}$. The calibration function is displayed in the legend.}
\end{figure}

\begin{figure}[t]
  \centerline{\includegraphics[width=\linewidth]{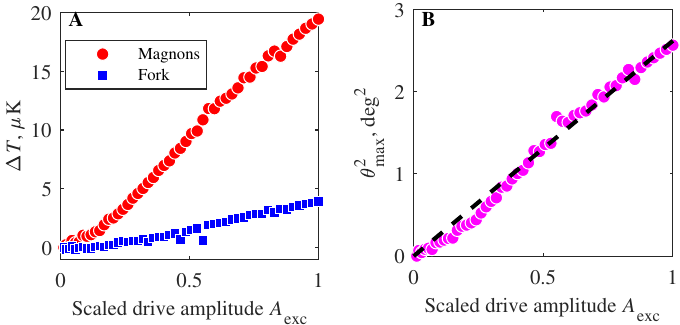}}
  \caption{\label{fig:SurfHeating} {\bf Tilt angle calibration.} ({\bf A}) Applied surface wave excitation leads to additional heating of the sample, measured both via relaxation rate of the bulk time crystal (red points), as well as via the thermometer fork (blue points). The observed temperature between the two location separated by 14$\,$cm is a proxy for dissipated power. ({\bf B}) The measured temperature difference is converted to tilt angle using Eq.~\eqref{eq:tiltcalib} (magenta points) and fitted with a single parameter linear fit resulting in $\theta_{\rm max}^2 {\rm (deg)}^2 \approx 2.6182\,A_{\rm exc}$ (black dashed line).}
\end{figure}

\clearpage

\renewcommand{\theequation}{S\arabic{equation}}

\section*{Supplemental Information}

\subsection*{Optomechanical coupling}

The optomechanical coupling term $g$ describes how the textural (order parameter) part of the trapping potential for magnons,  Eq.~(M17) in Methods, changes shape as the free surface of the superfluid is moving. Below we explain the different contributions to the textural free energy that compete and in combination create this effect. This allows for qualitative understanding of the coupling mechanisms. A quantitative description using a three-dimensional numerical simulation of the system is beyond the scope of the present Article.  

In the superfluid B phase studied here, the superfluid state is described by an order parameter $\mathcal{A}_{\alpha i} = \Delta \, e^{-i \phi} \, R_{\alpha i}$, where $\Delta$ is the superfluid gap, superflow arises as the gradient of the phase $\phi$, and $R_{\alpha i}$ is a three-dimensional rotation matrix that can be parametrised as  $R_{\alpha i} (\hat{\bf n},\theta)$. Here the rotation angle is fixed to $\theta\approx 104^\circ$ and the rotation axis $\hat{\bf n}$ is a unit vector field in space. The spatial variation of the order parameter relevant to the present work is entirely contained in $\hat{\bf n}$. The spatial distribution of $\hat{\bf n}$ results from the minimisation of the free energy of the order parameter configuration. Below we use the notation of Ref.~\citenum{thune_hydr} for these contributions.

\subsection*{Bulk time crystal}

The leading order free energy terms governing the texture in bulk fluid are the dipole-field term
\begin{equation} \label{eq:dipfield}
    F_{\rm DH} = - a \int d^3 r \left( \hat{\bf n} \cdot {\bf H} \right)^2 \,,
\end{equation}
the dipole-velocity term
\begin{equation} \label{eq:dipvel}
    F_{\rm DV} = - \lambda_{\rm DV} \int d^3 r \left[ \hat{\bf n} \cdot \left( {\bf v}_{\rm s} - {\bf v}_{\rm n} \right) \right]^2\,,
\end{equation}
the field-velocity term
\begin{equation} \label{eq:fieldvel}
    F_{\rm HV} = - \lambda_{\rm HV} H^2 \int d^3r \left[ \hat{\bf l} \cdot \left( {\bf v}_{\rm s} - {\bf v}_{\rm n} \right) \right]^2\,,
\end{equation}
the first-order field-velocity term
\begin{equation} \label{eq:1stfieldvel}
    F_{\rm HV1} = - \lambda_{\rm HV1} H \int d^3 r \left( \hat{\bf l} \cdot \nabla \times {\bf v}_{\rm n} \right) \,,
\end{equation}
and the gradient energy term
\begin{equation} \label{eq:gradterm}
    F_{\rm G} = \int d^3 r \left[ \lambda_{\rm G1} \frac{\partial R_{\alpha i}}{\partial r_i} \frac{\partial R_{\rm \alpha j}}{\partial r_j} + \lambda_{\rm G2} \frac{\partial R_{\alpha j}}{\partial r_j} \frac{\partial R_{\rm \alpha j}}{\partial r_i} \right]\,.
\end{equation}
The definitions and theoretical evaluation of the positive-valued pressure- and temperature-dependent parameters $a$, $\lambda_{\rm DV}$, $\lambda_{\rm HV}$, $\lambda_{\rm HV1}$, $\lambda_{\rm G1}$, and $\lambda_{\rm G2}$ can be found in Ref.~\citenum{thune_hydr}. The vector fields that appear above are the magnetic field $H=|{\bf H}|$, the superfluid and normal fluid flow fields ${\bf v}_{\rm s}$ and ${\bf v}_{\rm n}$, and the orbital anisotropy axis of the Cooper pairs $\hat{\bf l} = \frac{{\bf H}}{|{\bf H}|} \cdot R_{\alpha i} (\hat{\bf n},\theta)$. Note that the trapping potential in Eq.~(M17) in Methods) is given in terms of the tipping angle of the $\hat{\bf l}$ field, measured from the magnetic field direction.

In our typical experimental conditions ($T \approx 0.15 T_{\rm c}$, $H \gtrsim 200\,$G) the dipole-velocity term \eqref{eq:dipvel} is always a couple of orders of magnitude smaller than the field-velocity term \eqref{eq:fieldvel} and can be neglected. Similarly, we can neglect the first-order field-velocity term \eqref{eq:1stfieldvel} since we are working with a non-rotating sample (i.e. $\nabla \times {\bf v}_{\rm n} = 0$). Let us now estimate the magnitude of the remaining terms under these conditions.

Starting with the dipole-field term we have energy density $-F_{\rm DH}/V \sim a H^2 \approx 3 \cdot 10^{-9}\,$erg/cm$^3$ when $\hat{\bf n} \parallel {\bf H}$. This assumption is satisfied near the cylinder axis, that is, where the time crystals are located in the absence of a flow field because of the cylindrical symmetry. The energy density due to the field-velocity term is $-F_{\rm HV}/V \sim \lambda_{\rm HV} H^2 |v_{\rm s}|^2 \approx |v_{\rm s}|^2 \cdot 8 \cdot 10^{-8} \,$erg/(cm$\cdot$s$^2$) for $\hat{\bf l} \parallel {\bf v}_{\rm s}$. The terms are comparable, $F_{\rm HV} \sim F_{\rm F_{\rm DH}}$, for realistic velocities $|v_{\rm s}| \sim 1\,$mm/s. Finally, we can estimate the magnitude of the gradient term by noting that changes in the order parameter distribution occur typically across the magnetic healing length $\xi_{\rm H}$. Thus, $F_{\rm G} \sim (\lambda_{\rm G1} + \lambda_{\rm G2})/\xi_{\rm H}^2 \approx 10^{-9}\,$erg/cm$^3$.

From the above considerations we see that the relevant terms, the dipole-field term \eqref{eq:dipfield}, the field-velocity term \eqref{eq:fieldvel}, and the gradient energy term \eqref{eq:gradterm}, are comparable and therefore, all relevant. We should therefore expect that the optomechanical coupling $g$ between the moving surface and the bulk time crystal carries contributions from both static tilt of the free surface via the gradient term \eqref{eq:gradterm} and from the field-velocity term \eqref{eq:fieldvel}, which is absent for static tilt of the free surface. 

\subsection*{Surface time crystal}

Let us now estimate the order of magnitude of the relevant free energy terms near the free surface. The presence of a free surface gives rise to the surface-field term
\begin{equation} \label{eq:surffield}
    F_{\rm SH} = -d_{\rm SH} H^2 \int d^2 r \left( \hat{\bf l} \cdot \hat{\bf s} \right)^2\,,
\end{equation}
the first-order surface-field-velocity term
\begin{equation} \label{eq:1stsurffieldvel}
    F_{\rm SHV1} = - \lambda_{\rm SHV1} H \int d^2 r \left[ \hat{\bf l} \cdot \hat{\bf s} \times \left( {\bf v}_{\rm s} - {\bf v}_{\rm n} \right) \right]\,,
\end{equation}
the surface-gradient term
\begin{equation} \label{eq:surfgrad}
    F_{\rm SG} = \lambda_{\rm SG} \int d^2 r \hat{s}_j R_{\rm \alpha j} \frac{R_{\rm \alpha i}}{\partial r_i}\,,
\end{equation}
and the surface-dipole term
\begin{equation}
    F_{\rm SD} = \int d^2 r \left[ b_4 \left( \hat{\bf s} \cdot \hat{\bf n} \right)^4 - b_2 \left( \hat{\bf s} \cdot \hat{\bf n} \right)^2  \right]\,.
\end{equation}
Similarly to the bulk energy terms, the equations above serve as the definitions for $d_{\rm SH}$, $\lambda_{\rm SHV1}$, $\lambda_{\rm SG}$, $b_4$, and $b_2$. Here $\hat{\bf s}$ is a unit vector oriented perpendicular to the surface and pointing towards the fluid.

The surface-field term \eqref{eq:surffield} orients $\hat{\bf l}$ perpendicular to the surface. Its magnitude under the test conditions is $-F_{\rm SH}/A \approx 9\cdot 10^{-9}\,$erg/cm$^2$.

For velocity ${\bf v}_{\rm s}$ along the surface (setting ${\bf v}_{\rm n} = 0$), such as near $r = z = 0$ for the flow induced by the surface wave mode we are exciting, the first-order surface-field-velocity term \eqref{eq:1stsurffieldvel} works to orient $\hat{\bf l}$ along the surface but perpendicular to the flow. The prefactor obtains a non-zero value due to broken particle-hole symmetry and is given by \cite{thune_hydr}
\begin{equation}
    \lambda_{\rm SHV1} = \frac{4 m h_{\rm H1} \Delta^2 \xi_{\rm GL}^2}{h}\,,
\end{equation}
where
\begin{equation}
    g_{\rm H1} = \frac{1}{6} \gamma \hbar \left. \frac{\partial N}{\partial E} \right|_{E = E_{\rm F}} \ln \left( \frac{E_{\rm F}}{k_{\rm B} T_{\rm c}} \right)\,.
\end{equation}
We get an estimate $\lambda_{\rm SHV1} \approx 5 \cdot 10^{-13}\,$erg/(G$\cdot$cm$^2 \cdot$ cm/s), which results in $-F_{\rm SHV1}/A \approx 1 \cdot 10^{-10}\,$erg/(cm$^2$ $\cdot$ cm/s). 

The surface-gradient term \eqref{eq:surfgrad} acts to minimize the gradients along the surface. We can estimate its magnitude by using a similar substitution as for the bulk gradient term \eqref{eq:gradterm}, i.e. $\partial R/\partial r \sim 1/\xi_{\rm H}$. This results in $F_{\rm SG} \sim \lambda_{\rm SG}/\xi_{\rm H} \approx 2 \cdot 10^{-10}\,$erg/cm$^2$. 

Finally, let us estimate the surface-dipole term. According to Ref.~\citenum{thune_hydr} we have  $b_2 (P=0) \sim 17 g_{\rm D} \Delta ^2 \xi_{\rm GL}$ and $b_2 (P=0) \sim 5 g_{\rm D} \Delta ^2 \xi_{\rm GL}$, which result in a net-negative term $- F_{\rm SD}/A \sim 12 g_{\rm D} \Delta ^2 \xi_{\rm GL} \approx 1 \cdot 10^{-10}\,$erg/cm$^2$.

From the above considerations we see that for realistic velocities very close to the free surface, $v_{\rm s} \lesssim 1\,$cm/s, the surface-field term \eqref{eq:surffield} is approximately two orders of magnitude larger than the other surface terms, suggesting that the local texture is fixed by the surface orientation and not e.g. by the first-order surface-field-velocity term \eqref{eq:1stsurffieldvel}. That is, the optomechanical coupling $g$ for the surface time crystal should only consist of a static part that is the same regardless of whether the surface tilt is applied statically or dynamically.

\subsection*{Effect of magnons}

The time crystal frequency is seen changing in the experiments. This results from the magnons' contrtibution to the textural free energy via the spin-orbit interaction
\begin{equation} \label{eq:spinorbit}
    F_{\rm SO} = \int d^3 r \left[ \frac{4}{5} \hbar \frac{\Omega_{\rm L}^2}{\omega_{\rm L}} \sin ^2 \left( \frac{\beta_{\rm L}}{2} \right) \left| \Psi \right|^2 \right] \,,
\end{equation}
where the wave function is related to the magnon number as $|\Psi| \sim N_{\rm m}^{1/2}$. The term \eqref{eq:spinorbit} orients $\hat{\bf l}$ along the spin vector, i.e. along ${\bf H}$. Under our typical experimental conditions, $F_{\rm SO}/(V \cdot N_{\rm m}) \sim 1 \cdot 10^{-22} \,$erg/cm$^3$. This term becomes comparable to the bulk energy terms for $N_{\rm m} \gtrsim 10^{12}$, corresponding to the end of the signal where the fitted parameter values stop changing.  The time crystal phenomenology that results from this changing frequency is discussed in references \citenum{TimeCrystalJosephson,autti2022nonlinear}, the self trapping effect resulting from Eq.~(\ref{eq:spinorbit}) in references \citenum{self_trap,QBall,mitbaganalog}, the decay mechanisms that lead to the change in the magnon number in references \citenum{magrelax_JLTP175,thermom_JLTP175}, and spectral details of magnons in the trapping potential in references \citenum{zavjalov2015measurements,10.1063/5.0189649}.

\end{document}